# Optical Method for Determination of Carrier Density in Modulation Doped Quantum Wells

G.V.Astakhov[1,2], V.P.Kochereshko[1], D.R.Yakovlev[1,2], W.Ossau[2], J.Nürnberger[2], W.Faschinger[2], G.Landwehr[2], T.Wojtowicz[3], G.Karczewski[3], and J.Kossut[3]

[1]*A.F.Ioffe Physico-Technical Institute, Russian Academy of Sciences, 194021 St.Petersburg, Russia*

[2]*Physikalisches Institut der Universität Würzburg, D-97074 Würzburg, Germany*

[3]*Institute of Physics, Polish Academy of Sciences, PL-02608 Warsaw, Poland*

An optical method is suggested to determine the concentration of two-dimensional electrons in modulation-doped quantum wells at low and moderate electron densities between $10^9$ and $2\times 10^{11}$ cm$^{-2}$. The method is based on an analysis of magneto-reflectivity spectra of charged excitons (trions). The circular polarization degree and the oscillator strength of the charged excitons contain information about the density and spin polarization of two-dimensional electron gas. The method is applied to CdTe/(Cd,Mg)Te and ZnSe/(Zn,Mg)(S,Se) heterostructures.

PACS: 71.10 Ca, 71.35.-y, 73.20.Dx, 78.66.Hf

I. INTRODUCTION

Optical properties of low-dimensional semiconductor heterostructures with quantum wells (QWs) have been studied in a great detail for two limiting cases: (i) undoped QWs (containing no free carriers) where optical spectra near the band gap are dominated by excitons (that is correlated states of photogenerated electron-hole pairs), and (ii) QWs with high density of quasi two-dimensional electron or hole gas (2DEG or 2DHG, respectively) due to modulation doping in barriers. In the presence of a high density of carriers, the excitons are screened, but the electron-hole correlations caused by the Coulomb interaction show up in optical spectra as the, so-called, Fermi edge singularity (FES) [1].

Recently, the regime of low- and moderate densities of the free carriers in QWs became a subject of intensive investigation. It is characterized by the coexistence of excitons and excitonic complexes with free carriers. With an increasing density of the free carriers, excitonic optical spectra are modified and gradually transform into the FES regime [2-5]. An electron-exciton interaction induces a number of interesting effects characteristic for these low- and moderate carrier densities. Examples of those include formation of charged exciton states (trions) [6], combined exciton-cyclotron and trion-cyclotron resonances in magnetooptical spectra [7,8], shake-up processes in the emission spectra [9], and spin-dependent exciton-electron scattering contributions to a homogeneous linewidth of the excitons [10,11]. These studies were possible because of a high quality of II-VI QW structures, based on CdTe and ZnSe. The stronger Coulomb interaction in II-VI semiconductors, compared to that in III-V materials such as GaAs, makes II-VI heterostructures a very suitable model system for the optical studies of various phenomena induced by the exciton-electron interaction.

For such studies, a detailed information concerning the system of the free carriers, such as their concentration and degree of spin polarization by an external magnetic field, is of importance. One possible way to determine carrier concentration of low- and moderate densities is to use the method based on dimensional magnetoplasma resonance [12]. This method requires high mobility of the carriers and has been successfully applied for III-V heterostructures only. Unfortunately, application of usual magneto-transport methods or the cyclotron resonance technique, which were successfully used in the case of III-V heterostructures with highly mobile carriers [13], are very limited in wide-gap II-VI heterostructures because the latter exhibit relatively low mobility of electrons and holes. These methods can only be used in the case of modulation-doped QWs with high concentration of carriers exceeding $5\times10^{11}$ cm$^{-2}$ [14-17]. To investigate the regime of low- and moderate carrier densities, a reliable all-optical method of determination of the parameters of the free carrier systems is required.

It has been shown for II-VI heterostructures with high concentration of free carriers, when the excitonic effects are screened out, that the all-optical methods can be used to characterize the two-dimensional carrier gas. Sharp changes of the intensity and energy position of photoluminescence lines at integer filling factors [18,19] and the Moss-Burstein shift between the emission and absorption lines [20] are among them. In CdTe-based QWs the analysis of Moss-Burstein shift allows to measure the carrier concentrations only above $1.2\times10^{11}$ cm$^{-2}$. In diluted magnetic semiconductor QWs of (Cd,Mn)Te, due to a spin polarization induced by the 2DHG via the giant Zeeman splitting, the sensitivity of the method can be extended down to hole concentrations of $6\times10^{10}$ cm$^{-2}$ [20]. At high 2DEG (2DHG) densities, exceeding $10^{11}$ cm$^{-2}$, the carrier concentration in QWs can be also estimated from width of photoluminescence band, which is contributed by Fermi energy of carriers [21,22].

Another approach to the problem of assessing the properties of 2D carriers can makes use of the charged excitons (trions). Consisting of three carriers (two electrons and one hole in the case of the negatively charged exciton, and two holes and one electron in the case of the positively charged exciton), the trions formed by photon absorption have to incorporate a free carrier. As a result the oscillator strength of a trion optical transitions is directly linked to the 2D carrier density. At low densities these two quantities are directly proportional each other [11, 20].

In turn, the degree of spin polarization of the 2D carriers in external magnetic fields is also reflected by the polarization of the trions [23, 24].

In this paper we present an optical method to determine the carrier concentration in QW's at low- and moderate carrier densities (from $10^9$ cm$^{-2}$ to $2\times10^{11}$ cm$^{-2}$). The method is based on the analysis of the polarization of trions in magnetic fields and on the oscillator strength of trion resonances in reflectivity spectra. The method is demonstrated for CdTe/(Cd,Mg)Te and ZnSe/(Zn,Mg)(S,Se) quantum wells with 2DEG.

## II. SAMPLES AND EXPERIMENTAL

We have investigated modulation-doped CdTe/Cd$_{0.7}$Mg$_{0.3}$Te and ZnSe/Zn$_{0.89}$Mg$_{0.11}$S$_{0.18}$Se$_{0.82}$ single quantum well structures (SQW) grown by molecular-beam epitaxy on (100)-oriented GaAs substrates. The electron density in the QW's was varied from $n_e \approx 5\times10^9$ to $2\times10^{11}$ cm$^{-2}$. A band scheme for the studied CdTe- and ZnSe-based structures is presented in Fig.1.

CdTe/Cd$_{0.7}$Mg$_{0.3}$Te structures consist of an 80 Å wide CdTe SQW separated from the surface by a 750 Å wide Cd$_{0.7}$Mg$_{0.3}$Te barrier. The band offsets in the conduction band and in the valence band are $\Delta E_C \approx 370$ meV and $\Delta E_V \approx 155$ meV, respectively. The structures contain δ-doped layers with iodine impurities having concentration up to $2\times10^{17}$ cm$^{-3}$ positioned at a distance of $L = 100$ Å from the SQW (see Fig.1). A nominally undoped sample and three doped samples with various width of the δ-layer, namely 10, 20 and 50 Å wide, were studied. The samples with different electron densities were fabricated on the same substrate using the wedge growth mode. It allows to vary the doped δ-layer and to keep all the other relevant QW parameters (such as the QW width, the barrier height, background impurity concentration, etc.) constant at the same time (for details see Ref. [25]).

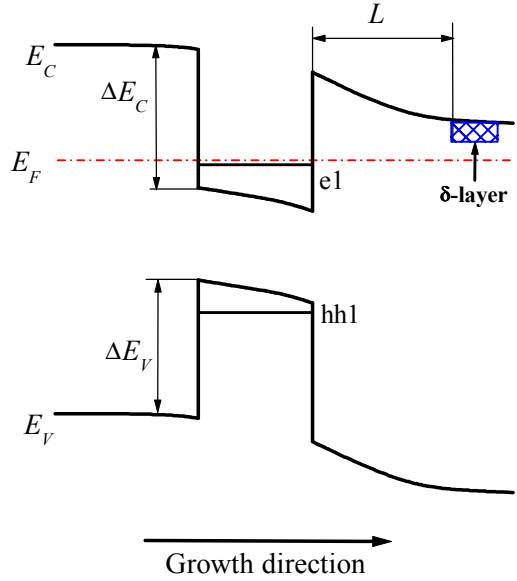

**Fig.1.** Band edge profiles in a heterostructure containing a single quantum well with a 2DEG and a δ-layer of donors.

The electron binding energy to an iodine shallow donor in CdTe is about 14 meV [26]. At low temperatures the electrons from the donors are moved over to the QW. As a result the δ-layer is positively charged while the QW is negatively charged. This induces an electric field and causes the band bending shown schematically in Fig.1. The estimated value of the electric field does not exceed $\Delta E_C/L \sim 10^5$ V/cm. Due to the large value of band discontinuity $\Delta E_C$ in comparison with the band bending in CdTe-based structures, most of the electrons from the donor δ-layer is trapped in the QW and the 2DEG density is nearly equal to the two-dimensional concentration of donors.

In ZnSe/Zn$_{0.89}$Mg$_{0.11}$S$_{0.18}$Se$_{0.82}$ structures an 80 Å wide ZnSe SQW is located between 1000 Å and 500 Å wide Zn$_{0.89}$Mg$_{0.11}$S$_{0.18}$Se$_{0.82}$ barriers. The structures contain 30 Å wide δ-layers at a distance of $L = 100$ Å from the QW doped with chlorine, whose donor binding energy in ZnSe is 26 meV [27]. A set of structures with different doping levels varying in the carrier concentration from $6\times10^{17}$ cm$^{-3}$ to $8\times10^{18}$ cm$^{-3}$ was fabricated. Since the band gap discontinuity in the conduction band is $\Delta E_C (\approx \Delta E_V) = 100$ meV, it could be compensated by the electric field due to the ionized donors in the δ-layer. As a result the Fermi level is pinned by the donors and only a small part of donor electrons (5-10%) is transferred to the QW.

Reflectivity spectra of the samples were measured at 1.6 K in external magnetic fields up to 7.5 T applied in the Faraday geometry. Achromatic quarter-wave length plates were used to analyze circularly polarized light. The light signal was detected by a charge-coupled device after being dispersed in a 1-m monochromator. A halogen lamp was used as a light source. Most of the reflectivity measurements were performed at normal incidence and only some spectra were taken under Brewster angle incidence in order to analyze transitions with very weak oscillator strength.

The carrier concentration in the modulation doped QWs is known to depend strongly on an additional, above-the-barrier illumination [20]. For low illumination intensity, the electron concentration in a QW increases with increasing intensity and, then, saturates at a high illumination power. The detailed analysis of this effect the structures studied is not a subject of this paper and will be discussed elsewhere [28]. Note that in our experiments we used the conditions of high intensity illumination controlled by a halogen lamp in order to achieve the maximum electron concentration in QWs.

III. MAGNETO-REFLECTIVITY SPECTRA OF TRIONS

Figure 2 shows the modification of reflectivity spectra in CdTe and ZnSe structures with increasing electron density. In the nominally undoped CdTe QW only the exciton resonance (X) at 1.634 eV is observable in the reflectivity spectrum displayed in Fig.2a. The presence of electrons in the QW with $n_e = 5 \times 10^9$ cm$^{-2}$ coming from the residual donors in the barrier layers has been detected by reflectivity measurements under the Brewster angle, the method described in the present paper in part VI. A new resonance of a negatively charged trion (T) appears at 1.630 eV in the spectra of doped structures. The amplitude of the trion line increases with the electron density growing and it dominates in the reflectivity spectrum for $n_e = 8\times10^{10}$ cm$^{-2}$. A broadening and a decrease of the amplitude of the excitonic resonance accompany this evolution. The observed modifications are typical and have been reported for CdTe/(Cd,Zn)Te QWs with 2DEG [3] and (Cd,Mn)Te/(Cd,Mg,Zn)Te QWs with 2DHG [20]. The evolution of the reflectivity spectra of ZnSe-based structures shown in Fig.2b is very similar to that in CdTe-based QWs. We note here that a detailed identification of the trion resonances in the reflectivity (absorption) spectra of modulation doped QWs have been published in Refs. [6,23].

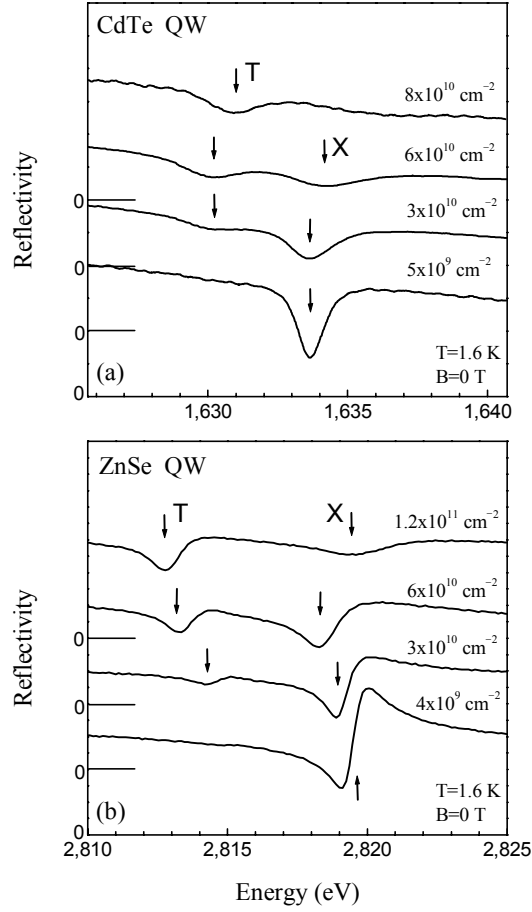

**Fig.2.** Reflectivity spectra of modulation doped SQWs with different electron density.
(a) 80 Å wide CdTe/Cd$_{0.7}$Mg$_{0.3}$Te SQWs. (b) 80 Å wide ZnSe/Zn$_{0.89}$Mg$_{0.11}$S$_{0.18}$Se$_{0.82}$ SQWs. The electron concentrations given in the figure were determined by an optical method introduced in the present work.

One of the distinct features of the trions is a strong circular polarization of their reflectivity (absorption) lines in the presence of a magnetic field [6]. In Fig.3a we present the reflectivity spectra of a CdTe-based sample with $n_e \approx 8\times10^{10}$ cm$^{-2}$ in a magnetic field $B$=7.5 T for two circular polarizations $\sigma^+$ and $\sigma^-$. Figure 3b shows the spectra of a ZnSe-based structure with $n_e \approx 6\times10^{10}$ cm$^{-2}$ in a magnetic field of 7.5 T. The strong polarization of the trion resonance observed for both samples is attributed to the fact that the trion ground state is a singlet. In a singlet state spins of two electrons forming the trion are antiparallel. In high enough magnetic fields, when the 2DEG is completely spin polarized, the trions can be created by a photon with one specific polarization only, namely in the polarization which photocreates an electron with the spin directed oppositely to the spins of the background electrons. It occurs for $\sigma^+$ polarization in CdTe-based structures with a negative electron g-factor [29], and for $\sigma^-$ polarization in ZnSe-based structures with a positive electron g-factor [30,31]. Some degree of polarization of the exciton reflectivity line can also be observed in these spectra. It is also caused by spin polarization of 2DEG, the details of the underlying mechanism desribed in Refs. [10,11,32].

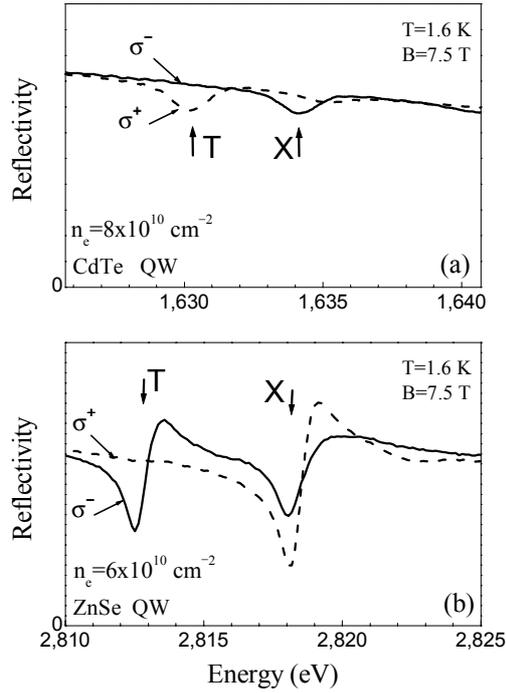

**Fig.3.** Reflectivity spectra measured in a magnetic field of 7.5 T in $\sigma^+$ and $\sigma^-$ circular polarization from: (a) 80 Å wide CdTe/Cd$_{0.7}$Mg$_{0.3}$Te SQW with $n_e = 8\times10^{10}$ cm$^{-2}$, and (b) 80 Å wide ZnSe/Zn$_{0.89}$Mg$_{0.11}$S$_{0.18}$Se$_{0.82}$ SQW with $n_e = 6\times10^{10}$ cm$^{-2}$.

The suggested method of determination of the 2DEG density exploits the fact that the polarization degree of the trion resonance $P_c^T(B)$, measured by means of the reflectivity, or absorption or transmission techniques, accurately renders the polarization of the 2DEG $P^{2DEG}(B) = P_c^T(B)$. The degree of the circular polarization of the trion reflectivity line can be determined from the experimentally measured quantities using the following relation:

$$P_c^T(B) = \frac{\Gamma_0^{T+} - \Gamma_0^{T-}}{\Gamma_0^{T+} + \Gamma_0^{T-}} \quad . \tag{1}$$

Here $\Gamma_0^{T+}$ and $\Gamma_0^{T-}$ are the oscillator strengths of the trion resonance detected in $\sigma^+$ and $\sigma^-$ circular polarizations, respectively. Oscillator strength of the exciton $\Gamma_0^X$ and the trion $\Gamma_0^T$ (i.e., the radiative damping) were extracted from the reflectivity spectra by the method described in Refs. [33,34] (see also Ref. [11]). The experimental dependencies $P_c^T(B)$ are plotted by circles in Figs. 4 and 5 for CdTe and ZnSe QWs with various electron concentrations. We will discuss in the next sections how the information on the 2DEG density can be extracted from these experimental data.

IV. SPIN POLARIZATION OF 2DEG

At low carrier densities and/or high temperatures, when the Fermi energy is smaller than the thermal energy $E_F < k_B T$, the spin polarization of the nondegenerate 2DEG is described by Boltzmann statistics

$$P^{2DEG}(B,T) = tanh\left(-\frac{\mu_B g_e B}{2k_B T}\right). \tag{2}$$

Here $\mu_B$ is the Bohr magneton, $k_B$ is the Boltzmann constant and $g_e$ is the g-factor of free electrons in the conduction band. At $T$=1.6 K ($k_B T = 0.14$ meV) the criterion $E_F < k_B T$ is met only for very dilute 2DEG with $n_e \leq 10^{10}$ cm$^{-2}$.

The Fermi-Dirac statistics should be used to describe the spin polarization of the 2DEG when $n_e > 10^{10}$ cm$^{-2}$ at $T$=1.6 K. The following procedure is used to calculate the spin polarization properties of a 2DEG, in such case

$$P^{2DEG}(B,T,n_e) = \frac{N^+ - N^-}{N^+ + N^-}. \tag{3}$$

Here $N^+$ ($N^-$) is the total number of electrons in all occupied Landau levels with a given spin orientation

$$N^{\pm}(E_F) = \sum_{i=0}^{\infty} N_i^{\pm}(E_F) \tag{4}$$

where $N_i^+$ ($N_i^-$) denotes the number of spin polarized electrons on the $i$-th Landau level, "+" and "-" label $+1/2$ and $-1/2$ Zeeman sublevels, respectively. Further we have

$$N_i^{\pm}(E_F) = \frac{eB}{h} \frac{1}{1 + \exp\left(\frac{\left(i+\frac{1}{2}\right)\hbar\omega_c \pm \frac{1}{2}g_e \mu_B B - E_F}{k_B T}\right)} \tag{5}$$

with $\frac{eB}{h}$ being the degeneracy of the Landau level, $\hbar\omega_c = \frac{\hbar eB}{m_e}$ is the cyclotron energy and $m_e$ is the conduction electron effective mass. $E_F$ is the Fermi energy, which depends on the electron concentration and the electron density of states. The latter quantity changes in an external magnetic field, making $E_F(B)$. In the absence of the field we have

$$E_F = \frac{\pi\hbar^2}{m_e} n_e \quad . \tag{6}$$

Usually, the electron density $n_e$ can be assumed to be independent of the magnetic field, and then

$$N^+(E_F) + N^-(E_F) = n_e = const(B) \quad . \tag{7}$$

To calculate $P^{2DEG}(B)$ one should first determine the $E_F(B)$ dependence for a fixed value of $n_e$ by inserting Eqs. (4,5) into Eq. (7). An example of the $E_F(B)$ dependence for ZnSe QWs with $n_e = 1.2 \times 10^{11}$ cm$^{-2}$ is plotted in Fig.6 for $T$=0 and 1.6 K. Then, substituting this dependence in Eq. (5), the spin polarization degree of 2DEG: $P^{2DEG}(B,T,n_e)$ can be calculated using Eq. (3). We stress here that the only adjustable parameter in this approach is the value of the electron density $n_e$. Therefore, by fitting Eq. (3) to the experimental dependencies $P_c^T(B)$ one can deduce the value of the electron density.

It can be shown that, in fact, quite substantial changes of the electron spin polarization occur for filling factors $v < 2$, when the lowest Landau level becomes partially occupied. In this case $\hbar\omega_c > E_F, k_BT$ and Eq. (3) can be simplified to

$$P^{2DEG}(B,T,n_e) = \frac{exp\left(-\frac{g_e\mu_B B}{2k_B T}\right) - exp\left(\frac{g_e\mu_B B}{2k_B T}\right)}{2\,exp\left(\frac{2E_F - \hbar\omega_c}{2k_B T}\right) + exp\left(-\frac{g_e\mu_B B}{2k_B T}\right) + exp\left(\frac{g_e\mu_B B}{2k_B T}\right)} \quad . \tag{8}$$

It is easy to see that Eq. (8) for $E_F \ll \hbar\omega_c$ it reduces to Eq. (2), i.e., to the case of the Boltzmann statistics. In the opposite limit of $E_F \gg \hbar\omega_c$ the lowest Landau level is fully occupied and $P^{2DEG} = 0$.

## V. DETERMINATION OF THE 2DEG CONCENTRATION FROM THE POLARIZATION DEPENDENCE OF THE TRION REFLECTIVITY

Let us recollect that our method of determination of the electron concentration uses the fact that the polarization of the trion line follows the polarization of 2DEG, i.e., that $P_c^T(B) = P^{2DEG}(B)$. These dependencies for the nondegenerate electron gas calculated with the use of Eq. (2) are plotted in Figs. 4 and 5 by the dashed lines. In the calculation we use $T$=1.6 K and electron g-factor values $g_e$=-1.46 for CdTe-based QWs [29] and $g_e$=+1.14 for ZnSe-based structures [30,31]. The trion polarization for the slightly doped samples ($n_e < 3\times10^{10}$ cm$^{-2}$) coincides very well with the dashed lines (not shown in the figures).

First we discuss the data for CdTe QWs. At a low electron concentration (Fig.4a) there is only a small deviation of the experimental data from the Brillouin function. This deviation increases for growing electron density (Fig.4b, c). At the highest electron concentrations (Fig.4c) the trion line is completely unpolarized in low magnetic fields up to 2 T and only in higher fields it begins to be polarized (for a qualitative behavior see limiting cases of Eq. (8)). The fitting curves for the Fermi-Dirac statistics [Eqs. (3-7)] of the experimentally measured polarization of the trions are shown in Fig. 4 by the solid lines. In these calculations we use the electron effective mass $m_e = 0.105\ m_0$ [17]. The electron concentration in the QW, $n_e = const(B)$ was, as mentioned above, the only fitting parameter. The best fit of the experimental data gives the following values of 2DEG density for three modulation doped QWs shown in Fig.4: $n_e = 3 \times 10^{10}$ cm$^{-2}$, $6 \times 10^{10}$ cm$^{-2}$ and $8 \times 10^{10}$ cm$^{-2}$. The accuracy of the method is estimated to be better than $1 \times 10^{10}$ cm$^{-2}$. This can be seen from Fig. 4b where calculations of $P^{2DEG}(B)$ for $n_e = 5 \times 10^{10}$ and $7 \times 10^{10}$ cm$^{-2}$ (dashed-dotted lines) are compared with the best fit curve for $n_e = 6 \times 10^{10}$ cm$^{-2}$ (solid line).

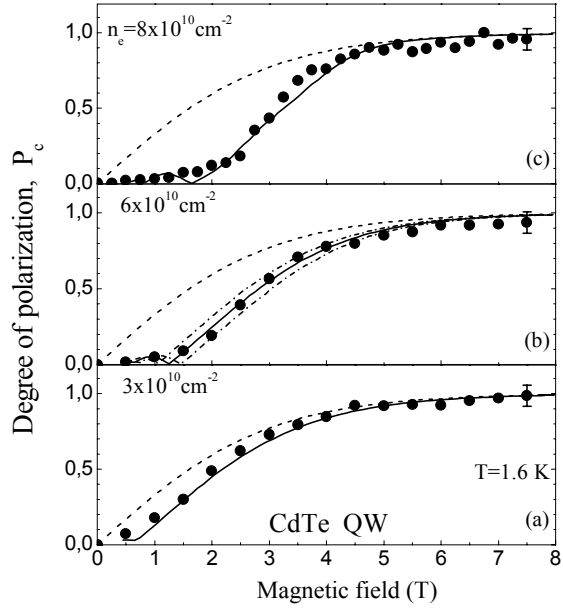

**Fig.4.** Magnetic-field-induced polarization of the trions in modulation doped CdTe/Cd$_{0.7}$Mg$_{0.3}$Te QWs. Experimental data (circles) are compared with model calculations for the nondegenerate- (dashed lines) and degenerate (solid and dashed-dotted lines) 2DEG. Electron densities determined from the best fit (solid lines) are given in the figure panels. Dashed-dotted lines in panel (b) correspond to calculation with $n_e = 5 \times 10^{10}$ and $7 \times 10^{10}$ cm$^{-2}$. They are given to show the accuracy of the method.

The same procedure has been performed for the doped ZnSe QWs shown in Fig.5. We use an electron effective mass of ZnSe $m_e = 0.15\ m_0$ [27] in these calculations. The solid lines show the results of the best fit subject to the condition $n_e = const(B)$. The electron density of $3 \times 10^{10}$ cm$^{-2}$, $6 \times 10^{10}$ cm$^{-2}$ and $1.2 \times 10^{11}$ cm$^{-2}$ are found for this set of samples. However, for the sample with highest $n_e$ the fitted curve deviates significantly from the experimental data points (see the solid line in Fig.5c) in the field range from 4 to 7 T, where $1 < \nu < 2$. We suggest this discrepancy be due to changes of the carrier concentration in the QW with an increasing magnetic field, i.e., that $n_e \neq const(B)$ in ZnSe QWs. In fact, in the case of the ZnSe-based QWs the band offset in the conduction band is not very large ($\Delta E_C \approx 100$ meV). Thus, for heavily doped structures the Fermi level can be pinned by the donors in the barrier, i.e. $E_F = const(B)$, that in turn will cause the oscillatory variation of the electron density in the QW with increasing magnetic field.

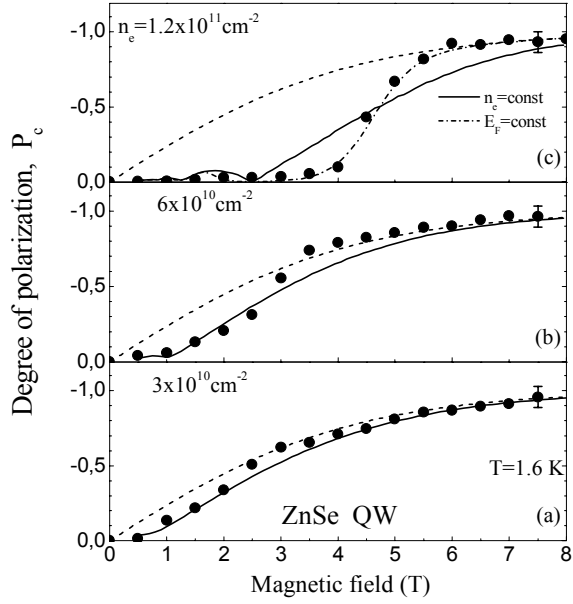

**Fig.5.** Magnetic-field-induced polarization of the trions in modulation doped ZnSe/Zn$_{0.89}$Mg$_{0.11}$S$_{0.18}$Se$_{0.82}$ QWs. Experimental data (circles) are compared with model calculations for the nondegenerate (dashed lines) and degenerate (solid lines) 2DEG. Electron densities determined from the best fit are given in the figure panels. The dashed-dotted line in the panel (c) shows the best fit for the degenerate 2DEG at $E_F = const(B)$.

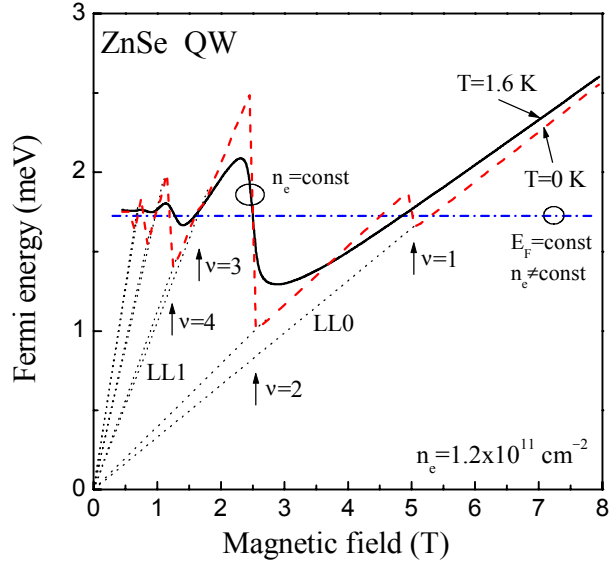

**Fig.6.** Fermi energy dependence on the magnetic field calculated in ZnSe QW with $n_e=1.2\times10^{11}$ cm$^{-2}$. The dashed line corresponds to the temperature equal to zero, the solid line is for $T$=1.6 K. In the calculations $n_e = const(B)$ was assumed. The dashed-dotted line shows the Fermi energy calculated under the assumption of $E_F = const(B)$. This assumption is realized in the case of the Fermi level pinned to the donors in the barriers. The dotted lines show the Landau levels fan.

The shift of the Fermi level induced by the magnetic field is shown in Fig.6. At zero temperature, assuming that the electron density does not depend on the magnetic field, the Fermi level follows the highest occupied Landau level and jumps to the lower Landau level at integer filling factors $\nu = n_e h/eB$ (see the dashed line). At a nonzero temperature (the solid line) the jumps of Fermi level become smooth. The position of the Fermi level at the vanishing magnetic field is shown by the dashed-dotted line in Fig.6.

The general solution for the 2DEG spin polarization by the magnetic field accounting for partial redistribution of electrons between barrier donors and the quantum well is rather complicated. Here we will analyze two limiting cases and will show that the choice of the modeling conditions gives minor corrections for evaluation of the carrier density. Let us consider two limiting cases: (i) the absence of the pining [$n_e = const(B)$] and (ii) the strong pining of Fermi level, $E_F = const(B)$. The first case is illustrated by the solid line in Fig.5c. In the second limit the polarization degree of 2DEG was obtained by means of Eqs. (3-5), with $E_F$ related to $n_e$ by Eq.(6) which is independent of the magnetic field. This fit, shown by the dashed-dotted line in Fig.5c, does describe the data points for the trion polarization degree. However, the determined value $n_e = 1.2 \times 10^{11}$ cm$^{-2}$ is with high accuracy the same for both these cases. Thus, as evident from Fig.5c, the polarization method is rather not sensitive to the choice of the conditions. It is seen also in this figure that two fitted dependencies calculated assuming either $n_e = const(B)$ or $E_F = const(B)$ cross at $P_c^T(B = 4.75 \text{ T}) \approx -0.5$. Therefore, as long as we are interested in the determination of $n_e$, the approach based on the condition $n_e = const(B)$ gives very reliable results. However, an additional care has to be taken when a detailed $P^{2DEG}(B)$ dependence for filling factors $1 < \nu < 2$ is required.

We should note here that, obviously, the polarization method could not be applied for the structures with $g_e \approx 0$, i.e. in conditions where the 2DEG polarization due to the thermal occupation of the Zeeman sublevels is vanishingly small. Our model calculations show that with the accuracy of $P_c$ determination better than $\pm 0.1$ the requirement for the method to be valid is $|g_e \mu_B B| > 0.2 k_B T$, for the range of magnetic fields corresponding to $\nu < 2$.

To summarize, the analysis of the circular polarization of the trion reflectivity line offers a reliable method to determine the 2DEG concentration. The low concentration limit of the method at $T = 1.6$ K is about $1 \times 10^{10}$ cm$^{-2}$. For lower concentrations 2DEG is nondegenerate, its polarization follows the Brillouin function and is insensitive to $n_e$. The high concentration limitation of the method is about $2 \times 10^{11}$ cm$^{-2}$, i.e., when the distinct trion line vanishes from the reflectivity spectra.

## VI. DETERMINATION OF 2DEG CONCENTRATION FROM TRION OSCILLATOR STRENGTH

It has been shown recently that in the region of low electron concentrations the oscillator strength of the trion resonance increases linearly with the electron density. In ZnSe-based QWs the linear dependence is valid for $n_e \leq 5 \times 10^{10}$ cm$^{-2}$ [11]. Once calibrated for the certain structures, the dependence of the trion oscillator strength on $n_e$ can be used for determination of electron density. We will show in this section that a sensitivity of this method allows us to determine the parameters characterizing very dilute electron gas with $n_e \approx 5 \times 10^9$ cm$^{-2}$.

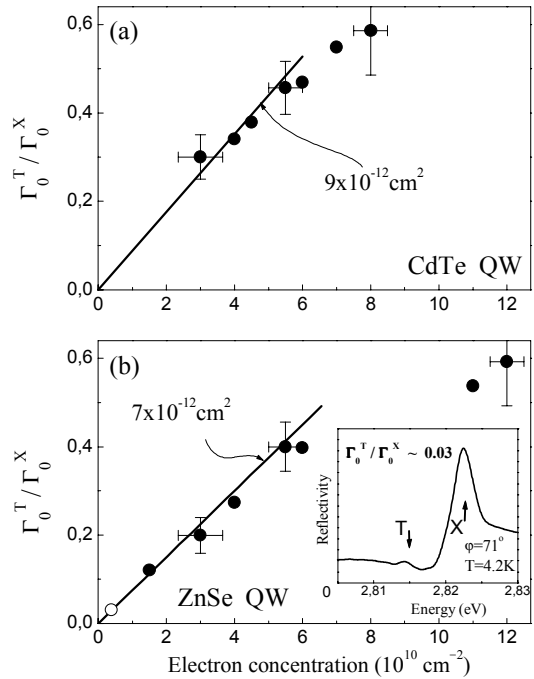

**Fig.7.** Dependence of the ratio of the trion oscillator strength and that of the exciton on electron density at $T$=1.6 K. (a) 80 Å wide CdTe/Cd$_{0.7}$Mg$_{0.3}$Te QWs. (b) 80 Å wide ZnSe/Zn$_{0.89}$Mg$_{0.11}$S$_{0.18}$Se$_{0.82}$ QWs. The inset shows the reflectivity spectrum measured at $T$=4.2 K at the oblique incidence φ=71° from a nominally undoped sample. The evaluated electron density is $n_e = 5 \times 10^9$ cm$^{-2}$. The open circle shows this value in the main panel.

At $n_e \leq 5 \times 10^{10}$ cm$^{-2}$ both the trion and the exciton resonances are very pronounced in the reflectivity spectra of II-VI QWs (see Fig.2 and Ref. [11]). In this case it is very convenient to deal with the trion oscillator strength normalized to that of the exciton, $\Gamma_0^T/\Gamma_0^X$. Moreover, for $n_e \leq 3\times10^{10}$ cm$^{-2}$, when a contribution of the electron-exciton scattering is negligibly small and the exciton and the trion resonances have about the same linewidth due to the inhomogeneous broadening, one does not have to extract the absolute values of $\Gamma_0^T$ and $\Gamma_0^X$ from the reflectivity spectra. In such a case the value of $\Gamma_0^T/\Gamma_0^X$ can be evaluated with a good accuracy from the ratio of the amplitudes of the trion and exciton resonances. In this method a slope $C$ of a linear dependence

$$\frac{\Gamma_0^T}{\Gamma_0^X} = C\, n_e \quad , \qquad (9)$$

has to be measured.

The experimental dependencies $\Gamma_0^T/\Gamma_0^X$ on $n_e$ for CdTe- and ZnSe-based QWs are plotted in Fig.7. For the data points shown by closed circles we determined $n_e$ via the polarization of the trions, the method described in section V. For both types of QWs a deviation from the linear dependence, shown by the solid lines, occurs for $n_e > 5\times10^{10}$ cm$^{-2}$. The coefficient $C$ in 80 Å wide CdTe/Cd$_{0.7}$Mg$_{0.3}$Te QW is $C=9\times10^{-12}$ cm$^2$ and in 80 Å wide ZnSe/Zn$_{0.89}$Mg$_{0.11}$S$_{0.18}$Se$_{0.82}$ QW we obtained $C=7\times10^{-12}$ cm$^2$ [35]. The value of $C$ depends on the trion area [11] and, therefore, it is expected to decrease in QWs with larger binding energies of the trions having more compact states. The relative accuracy of this method, which is directly linked with the accuracy of determination the coefficient $C$ shown by arrows bar in Fig.7, is better than 20% of $n_e$.

The sensitivity of the method discussed in this section depends on the accurate determination of $\Gamma_0^T/\Gamma_0^X$. In the reflectivity spectra of II-VI QWs measured at a normal incidence the trion resonance became very weak for $n_e$ decreasing to $1\times10^{10}$ cm$^{-2}$. Further increase of the sensitivity can be achieved by detecting the reflectivity spectra at an oblique light incidence at angles close to the Brewster angle [36,34] or by means of the modulation spectroscopy [37]. In the inset to Fig.7b we show a reflectivity spectrum taken from the nominally undoped ZnSe-based structure at an oblique incidence of light, at $\varphi = 71^o$ close to the Brewster angle ($\varphi_{Br} = 69^o$). From this spectrum we have deduced the ratio $\Gamma_0^T/\Gamma_0^X \approx 0.03$, and then with use of Eq. (9) we determined the electron concentration in the QW as $n_e \approx 4\times10^9$ cm$^{-2}$. The open circle in Fig.7b marks this data point. We estimate that the potential of this method to evaluate electron densities is as small as $10^9$ cm$^{-2}$.

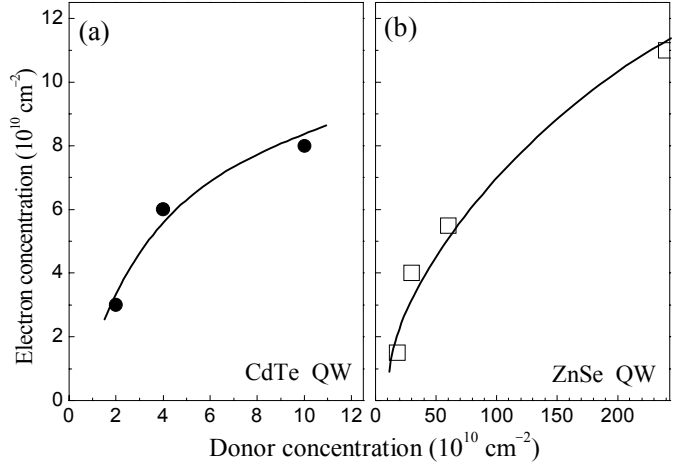

**Fig.8.** The electron density in the QWs as a function of the donor concentration in δ-layers determined for (a) CdTe/Cd$_{0.7}$Mg$_{0.3}$Te QWs and for (b) ZnSe/Zn$_{0.89}$Mg$_{0.11}$S$_{0.18}$Se$_{0.82}$ QWs at $T$=1.6 K.

VII. DISCUSSION

Figure 8 summarizes the determined electron densities in the studied structures by displaying them as functions of the donor concentration in the $\delta$-layer. One should keep in mind that there is a relatively large uncertainty of the donor concentration values which are based on a technological calibration of growth regimes. With this reservation we can conclude that in CdTe-based QWs all electrons from the donors are moved over to the QW. In contrast, in ZnSe-based QWs only a small portion of the donors (5-10%) transfer their electrons to the QW.

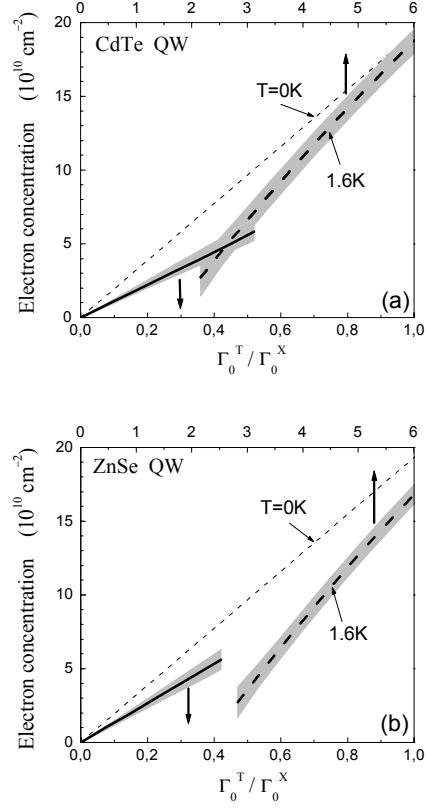

**Fig.9**. Calibration of $n_e$ determination using various optical methods described in the present paper. Electron density is related to $\Gamma_0^T/\Gamma_0^X$ by linear dependencies (solid lines) and to the magnetic field value $B_{1/2}$ (thick and thin dashed lines for $T$=1.6 and 0 K, respectively): (a) CdTe/Cd$_{0.7}$Mg$_{0.3}$Te QWs; (b) ZnSe/Zn$_{0.89}$Mg$_{0.11}$S$_{0.18}$Se$_{0.82}$ QWs. Gray colored areas represent accuracy of the electron density determination.

We have mentioned already by presenting the results of Fig.5c that the polarization method is rather insensitive to the choice of the conditions. The two fitted curves calculated under the condition $n_e = const(B)$ or $E_F = const(B)$ cross each other at $P_c^T(B = 4.75 \text{ T}) \approx -0.5$. With this result in mind we may conclude that the value of a magnetic field $B_{1/2}$ at which $\left|P_c^T(B_{1/2})\right| = 0.5$ can be used for evaluation of the electron density for both types of conditions. The dependence $n_e(B_{1/2})$ calculated for $T$=0 and 1.6 K is shown in Fig.9 by thin and thick dashed lines, respectively. One can see in Fig.9 that the sensitivity of the polarization method increases with decreasing temperature.

Figure 9 in fact summarizes the main results of the optical methods suggested in the present paper. The applicability ranges of the methods are shown. The analysis of the trion polarization at $T$=1.6 K is reliable in the concentration range from $3\times10^{10}$ cm$^{-2}$ to $2\times10^{11}$ cm$^{-2}$, which corresponds to the thick dashed line. We believe that the sensitivity in the low electron density range may be improved by taking data at a lowest possible temperature. Thin dashed line traces the limiting case of $T$=0 K.

Solid lines in Fig.9 display the part of the dependence $\Gamma_0^T/\Gamma_0^X = Cn_e$ where it can be interpolated by the linear relationship. One can see that both for CdTe- and ZnSe-based QWs linear relation becomes invalid for the electron density exceeding $5.5\times10^{10}$ cm$^{-2}$. The lower limit for the method based on the analysis of the trion oscillator strength depends on the sensitivity of the optical techniques used for evaluation of the oscillator strength. In II-VI QWs the reflectivity measurements at oblique incidence of the light allow to detect $2\times10^9$ cm$^{-2}$ electrons.

Gray colored areas in Fig.9 represent the accuracy of the used methods. For the polarization method the accuracy is $1\times10^{10}$ cm$^{-2}$. For the method based on the oscillator strength the relative accuracy is better than 20% of the electron concentration value.

In this paper we have limited ourselves to the optical methods which are suitable for relatively low electron densities. For electron densities exceeding $2\times10^{11}$ cm$^{-2}$ optical spectroscopy provides several other possibilities. Among them are: (i) analysis of the Moss-Burstein shift between the emission and absorption lines [20], and (ii) sharp changes of the photoluminescence intensity and of the energy position of the excitonic lines at integer filling factors [18,19].

In conclusion, reflectivity spectra taken from modulation doped CdTe/(Cd,Mg)Te and ZnSe/(Zn,Mg)(S,Se) QW structures in magnetic fields have been analyzed. An optical method of measuring 2DEG density is proposed. The method is based on the analysis of the circular polarization degree of the trion resonance and on the linear dependence of the trion oscillator strength on the electron density. The method could be used in the range of concentrations from $10^9$ cm$^{-2}$ up to $10^{11}$ cm$^{-2}$. We have demonstrated the efficiency of this method using quantum well structures based on CdTe as well as ZnSe with different barrier heights.

## ACKNOWLEDGMENTS

This work was supported by grants of the Russian Foundation for Basic Research (Grants Nos. 00-02-04020 and 01-02-16990) and the Deutsche Forschungsgemeinshaft (Grant No. Os98/6 and SFB 410), as well as by the Program "Nanostructures" of Russian Ministry of Science. We feel deep sorrow at the death of our colleague W. Faschinger, who was an outstanding scientists and very kind person.

# Literature

[1] J.C. Maan, in *The Physics of Low-Dimensional Semiconductor Structures*, edited by P.N.Butcher *et al*. (Plenum Press, New York, 1993), p.333.

[2] G. Finkelstein, H. Shtrikman, and I. Bar-Joseph, Phys. Rev. Lett. **74**, 976 (1995).

[3] V. Huard, R.T. Cox, K. Saminadayar, A. Arnoult, and S. Tatarenko, Phys.Rev.Lett. **84**, 187 (2000).

[4] J.A. Brum and P. Hawrylak, Comments Condens. Matter Phys. **18**, 135 (1997).

[5] E.I. Rashba and M.D. Sturge, Phys. Rev. B **63**, 045305 (2000).

[6] K. Kheng, R.T. Cox, Y. Merle d'Aubigne, F. Bassani, K. Saminadayar and S. Tatarenko, Phys. Rev. Lett. **71**, 1752 (1993).

[7] D.R. Yakovlev, V.P. Kochereshko, R.A. Suris, H. Schenk, W. Ossau, A. Waag, G. Landwehr, P.C.M. Christianen, and J.C. Maan, Phys.Rev.Lett. **79**, 3974 (1997).

[8] V.P. Kochereshko, D.R. Yakovlev, G.V. Astakhov, R.A. Suris, J. Nürnberger, W. Faschinger, W. Ossau, G. Landwehr, T. Wojtowicz, G. Karczewski, and J. Kossut, in *Optical properties of Semiconductor Nanostructures*: NATO Science Series, 3. High Technology – Vol. 81, Eds. M.L.Sadowski et al., 2000. pp.299-308.

W.Ossau, V.P.Kochereshko, G.V.Astakhov, D.R.Yakovlev, G.Landwehr, T.Wojtowicz, G.Karczewski, and J.Kossut, Physica B, *in press* (2001).

[9] G. Finkelstein, H. Shtrikman, and I. Bar–Joseph, Phys. Rev. B **53**, 12593 (1996).

[10] V.P. Kochereshko, D.R. Yakovlev, A.V. Platonov, W. Ossau, A. Waag, G. Landwehr, and R. Cox, in Proc. 23 Int. Conf. Physics of Semiconductors, Berlin, Germany 1996, (World Scientific, Singapore 1996), eds. M. Scheffler and R. Zimmermann, p.1943.

[11] G.V. Astakhov, V.P. Kochereshko, D.R. Yakovlev, W. Ossau, J. Nürnberger, W. Faschinger, G. Landwehr, Phys.Rev.B **62**, 10345 (2000).

[12] S.I. Gubarev, I.V.Kukushkin, S.V. Tovstonog, M.Yu. Akimov, J. Smet, K. von Klitzing, and W. Wegscheider, JETP Letters **72**, 324 (2000). [Pis'ma v Zhurnal Eksperimental'noĭ i Teoreticheskoĭ Fiziki **72**, 469 (2000)].

[13] I.Kukushkin and V.B.Timofeev, Adv. Phys. **45**, 147 (1996).

[14] S. Scholl, H. Schäfer, A. Waag, D. Hommel, K. von Schierstedt, B. Kuhn-Heinrich, and G.Landwehr, Appl. Phys. Lett. **62**, 3010 (1993).

[15] I.P. Smorchkova, N. Samarth, J.M. Kikkawa, and D.D. Awschalom, Phys. Rev. Lett. **78**, 3571 (1997).

[16] G. Karczewski, J. Jaroszynski, A. Barcz, M. Kutrowski, T. Wojtowicz, and J. Kossut, J. Crystal Growth **184/185**, 814 (1998).

[17] Y. Imanaka, T. Takamasu, G. Kido, G. Karczewski, T.Wojtowicz, and J.Kossut, Physica B **256-258**, 457 (1998).

[18] Y. Imanaka, T. Takamasu, G. Kido, G. Karczewski, T.Wojtowicz, and J.Kossut, J.Crystal Growth **214/215**, 240 (2000).

[19] M. Potemski, Physica B **256-258**, 283 (1998).

[20] P. Kossacki, J. Cibert, D. Ferrand, Y. Merle d'Aubigne, A. Arnoult, A. Wasiela, S. Tatarenko, and J. Gaj, Phys. Rev. B **60**, 16018 (1999).

[21] A.J. Shields, J.L. Osborne, M.Y. Simmons, D.A. Ritchie, and M. Pepper, Semicond. Sci. Technol. **11**, 890 (1996).

[22] V.P. Kochereshko, G.V. Astakhov, D.R. Yakovlev, W. Ossau, G. Landwehr, T. Wojtowicz, G. Karczewski, and J. Kossut, phys. Stat. Sol. (b) **220**, 342 (2000).

[23] G.V. Astakhov, D.R. Yakovlev, V.P. Kochereshko, W. Ossau, J. Nürnberger, W. Faschinger, and G. Landwehr, Phys. Rev. B **60**, R8485 (1999).

[24] R.T. Cox, V. Huard, K. Kheng, S. Lovisa, R.B. Miller, K. Saminadayar, A. Arnoult, J. Cibert, S. Tatarenko, and M. Potemski, Acta Physica Polonica A **94**, 99 (1998).

[25] T. Wojtowicz, M. Kutrowski, G. Karczewski, J. Kossut, Acta Physica Polonica A **94**, 199 (1998).

[26] J.M. Francou, K. Saminadayar, and J.L. Pautrat, Phys. Rev. B **41**, 12035 (1990).

[27] Landolt-Börnstein, New Series Vol. III/41B, "II-VI and I-VI Compounds; Semimagnetic Compounds", ed. by U.Rössler (Springer, Berlin, Heidelberg 1999).

[28] G.V. Astakhov, D.R. Yakovlev, V.P. Kochereshko, to be published.

[29] A.A. Sirenko, T. Ruf, M. Cardona, D.R. Yakovlev, W. Ossau, A. Waag, G. Landwehr, Phys. Rev. B **56**, 2114 (1997).

[30] D. R. Yakovlev, H. A. Nickel, B. D. McCombe, A. Keller, G. V. Astakhov, V.P. Kochereshko, W. Ossau, J. Nürnberger, W. Faschinger, G. Landwehr, J. Crystal Growth **214-215**, 823 (2000).

[31] J.J. Davies, D. Wolverson, I.J. Griffin, O.Z. Karimov, C.L. Orange, D. Hommel, M. Behringer, Phys. Rev. B. **62**, 10329, (2000).

[32] V.P. Kochereshko, A.V. Platonov, F. Bassani, and R. Cox, Supelatt.&Microstruct. **24**, 269 (1997).

[33] E.L. Ivchenko, A.V. Kavokin, V.P. Kochereshko, G.R. Posina, I.N. Uraltsev, D.R. Yakovlev, R.N. Bicknell-Tassius, A. Waag, G. Landwehr, Phys. Rev. B. **46**, 7713 (1992).

[34] E.L. Ivchenko and G.E. Pikus, In: *Superlattices and Other Heterostructures*. Springer-Verlag, Berlin, 1997.

[35] The value of *C* depends on the structure parameters (choice of semiconductor material and confinement conditions of the carriers). We have found that in ZnSe/Zn$_{0.86}$Be$_{0.06}$Mg$_{0.06}$Se QWs, whose band parameters are nearly identical to those in ZnSe/Zn$_{0.89}$Mg$_{0.11}$S$_{0.18}$Se$_{0.82}$ QWs, *C* value depend approximately linearly on the trion binding energy in the range 1.4 - 6.6 meV. It decreases from $9\times10^{-12}$ cm$^2$ in a 200 Å wide QW with the trion binding energy of 1.4 meV to $5.5\times10^{-12}$ cm$^2$ in a 50 Å wide QW with the trion binding energy of 6.6 meV. These results will be published elsewhere.

[36] E.L. Ivchenko, P.S. Kopev, V.P. Kochereshko, I.N. Uraltsev, D.R. Yakovlev, S.V. Ivanov, B.Ya. Meltser, and M.A. Kalitievskii, Sov. Phys. Semicond. **22**, 495 (1988). [Fiz. Tekh. Poluprovodn. **22**, 784 (1988)].

[37] M. Cardona, *Modulation Spectroscopy*. Solid State Physics, Suppl. 11. Eds. F. Seitz, D. Turnbull, and H. Ehrenreich. Academic Press, N.-Y., 1969.